\documentclass[journal,twoside,web]{ieeecolor}
\usepackage{tmi}
\usepackage{cite}
\usepackage{amsmath,amssymb,amsfonts}
\usepackage{algorithmic}
\usepackage{graphicx}
\usepackage{textcomp}
\usepackage{booktabs}
\usepackage{epstopdf}

\usepackage[colorlinks=true, allcolors=ieeeblue]{hyperref}
\usepackage{tikz}
\usetikzlibrary{calc}
\definecolor{ieeeblue}{RGB}{0,102,153}

\def\BibTeX{{\rm B\kern-.05em{\sc i\kern-.025em b}\kern-.08em
    T\kern-.1667em\lower.7ex\hbox{E}\kern-.125emX}}
\begin{document}

\title{A Pre-trained EEG-to-MEG Generative Framework for Enhancing BCI Decoding}
\author{Zhuo Li, Shuqiang Wang
\thanks{Zhuo Li and Shuqiang Wang are with Shenzhen Institutes of Advanced Technology, Chinese Academy of Sciences, Shenzhen 518055, China, and also with University of Chinese Academy of Sciences, Beijing 100049, China}
}

\maketitle

\begin{tikzpicture}[remember picture, overlay]
    \fill[white] (current page.north west) rectangle ($(current page.north east)+(0,-1.7cm)$);
    \coordinate (header_center) at ($(current page.north)+(0,-1.7cm)$);

    \draw[color=ieeeblue, line width=1pt]
         ($(header_center) - (0.5\textwidth, 0)$) -- ($(header_center) + (0.5\textwidth, 0)$);

    \node[anchor=south east, inner sep=0pt, yshift=3pt, font=\sffamily]
         at ($(header_center) + (0.5\textwidth, 0)$) {\thepage};
\end{tikzpicture}

\begin{abstract}
Electroencephalography (EEG) and magnetoencephalography (MEG) play important and complementary roles in non-invasive brain-computer interface (BCI) decoding. However, compared to the low cost and portability of EEG, MEG is more expensive and less portable, which severely limits the practical application of MEG in BCI systems. To overcome this limitation, this study proposes the first cross-modal generation framework based on EEG-MEG spatiotemporal coupled representations to synthesize MEG signals cost-effectively. The framework first extracts general neural activity representations through a pre-trained EEG model. Building upon these representations, the framework effectively learns the lower spatial dispersion and higher high-frequency sensitivity of MEG via the spatial focus mapping module and the broadband spectral calibration module. Experimental results demonstrate that the synthesized MEG signals show high consistency with the real MEG in both time-frequency characteristics and source space activation patterns. More importantly, downstream BCI decoding experiments demonstrate that using synthesized MEG leads to performance enhancements not only on paired EEG-MEG datasets but also on independent EEG-only datasets. Overall, this framework opens a new avenue for overcoming data bottlenecks in BCI.
\end{abstract}


\section{Introduction}
\label{sec:introduction}
\begin{figure}[ht]
\centering
\includegraphics[width=\linewidth]{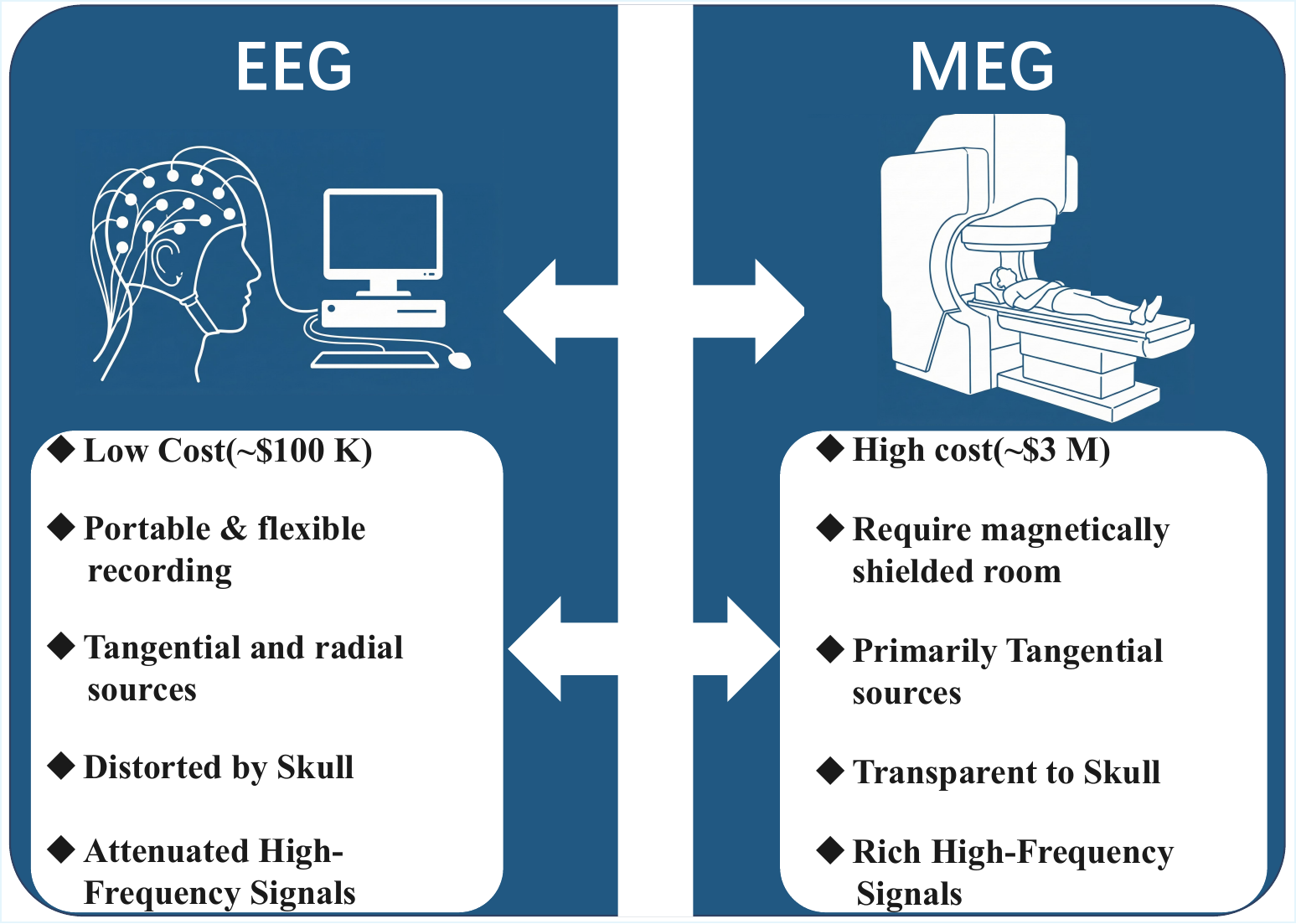}
\caption{Comparative analysis of the strengths and limitations of EEG and MEG.}
\label{fig:compare}
\end{figure}

\IEEEPARstart{I}{n} the field of non-invasive BCI, EEG and MEG have attracted significant attention due to their ability to capture neural activity in real-time\cite{guo2026generative, edelman2024non, wang2026non}. Nevertheless, as shown in Fig.~\ref{fig:compare}, EEG and MEG show distinct differences and complementarities in terms of practical deployment and neural representation. Benefiting from the fact that head tissues are essentially transparent to magnetic fields, MEG precisely captures tangential current sources and high-frequency neural oscillations\cite{fabbri2025noninvasive, vorwerk2025global, feys2025scalp}. Therefore, MEG provides detailed information on local neural dynamics. Conversely, while limited by volume conduction\cite{morik2025enhancing, nunez2007study}, EEG has a wider sensitivity range. EEG effectively captures radial current sources that are difficult for MEG to detect, providing information complementary to MEG\cite{feng2025explaining, geller2024magnetoencephalography}. From the standpoint of neural manifolds\cite{perich2025neural}, EEG and MEG represent projections of the high-dimensional neural state space onto magnetic and electric subspaces, respectively. By fusing complementary information from EEG and MEG, the limitations of neural manifold representations from single-modal views are mitigated\cite{li2025brainflora}, thereby overcoming performance bottlenecks in BCI decoding\cite{ji2025enhanced, bore2021long, jiang2024evaluation, li2023hybrid}.

However, the simultaneous acquisition of EEG and MEG signals requires expensive and complex hardware\cite{seedat2024simultaneous}. Compared to portable and low-cost EEG, MEG is expensive and needs a strict magnetic shielding environment. This severely limits its application in BCI systems. Furthermore, MEG cannot be used for individuals with metal implants, such as pacemakers or deep brain stimulators, which further limits its user population. These practical constraints make it difficult to acquire paired EEG-MEG data in real-world scenarios\cite{schwartz2025potential, schofield2024novel}. Consequently, it is difficult for current BCI systems to leverage fused information to enhance decoding performance.

At a more fundamental level, EEG and MEG are instantaneous and synchronous manifestations of the same neural current sources in electric and magnetic fields\cite{cohen1976part, cohen1983demonstration, sanchez2024solving}. This physical commonality establishes the spatiotemporal coupling between EEG and MEG, exhibiting high spatial correlation and zero-lag temporal alignment\cite{lankinen2024neuronal, jin2024bayesian, cho2024comparison}. This physical coupling provides the theoretical foundation for leveraging generative AI\cite{zong2024new, yao2025catd, li2025scdm} to model EEG-MEG mappings within the shared latent source space, thereby enabling the cost-effective synthesis of MEG signals.

In practice, strict recording requirements severely limit the amount of paired EEG-MEG data. This lack of data makes it very difficult to train complex EEG-to-MEG generative models from scratch, as models are highly prone to overfitting\cite{liang2026generative, duan2024retain}. To reduce the dependence on large-scale paired data, pre-trained models trained on massive unlabeled datasets offer a promising solution. Through self-supervised learning, these pre-trained models capture complex spatiotemporal dependencies from large-scale neural recordings\cite{10561479, wang2024eegpt}. For instance, Cui et al. \cite{cui2024neuro} validate the effectiveness of foundation models in EEG, demonstrating that large-scale pre-training significantly mitigates the data scarcity issue in downstream tasks. Similarly, Jiang et al. \cite{jiang2024large} propose a unified framework compatible with varying electrode configurations, successfully capturing generic neural representations from massive diverse recordings. This pre-training approach effectively addresses distribution shifts across subjects and provides robust prior knowledge for downstream tasks. Therefore, leveraging pre-trained models to use their general representations is an effective way to reduce the need for large paired datasets and improve model performance\cite{11204282}.

In this study, a cross-modal generation framework based on EEG-MEG spatiotemporal coupled representations is proposed to synthesize MEG. The pre-trained model is utilized as the backbone to acquire general neural activity representations. Upon this foundation, specialized modules are designed to fully leverage the EEG-MEG spatiotemporal coupling, ensuring that the synthesized MEG preserves the low spatial dispersion and broadband characteristics inherent to real MEG. Results demonstrate that combining the original EEG with the synthesized MEG effectively enhances BCI decoding performance, even for subjects with only EEG data available. The primary contributions of this study are threefold:

\begin{enumerate}
    \item A paradigm based on a pre-trained EEG model is proposed for synthesizing MEG from EEG. By leveraging synthetic MEG, the framework expands the observation space within the neural state space and supplements the information missing in EEG projections on the neural manifold. To our knowledge, this is the first time a cross-modal generation framework based on EEG-MEG spatiotemporal coupled representations has been developed for cost-effective synthesis of MEG signals.
   \item A spatial focus mapping module is designed to model the spatial coupling between EEG and MEG. Addressing the projection differences of shared neural sources between electric and magnetic field projections, this module employs spatiotemporal attention and vector quantization to map the diffuse spatial representations of EEG onto the focused spatial distributions specific to MEG.
   \item A broadband spectral calibration module is designed to recover the inherent broadband spectral characteristics of MEG. By utilizing a learnable global frequency filter and multi-band supervision, this module ensures that the synthesized MEG is consistent with real MEG across both low-frequency and high-frequency neural oscillations.
\end{enumerate}

\section{Method}

\subsection{Framework Overview}
This study aims to leverage EEG-MEG spatiotemporal coupled representations to synthesize MEG signals cost-effectively. Let $X=\{X_i\}_{i=1}^N$ denote the set of observed EEG time series, and $Y=\{Y_i\}_{i=1}^N$ be the corresponding synchronized MEG signals. A single sample $X \in \mathbb{R}^{C_e \times T}$ contains $C_e$ electrode channels and $T$ time points, while the target signal $Y \in \mathbb{R}^{C_m \times T}$ consists of $C_m$ sensor channels. Our goal is to learn a parameterized mapping function $\mathcal{F}_\theta$. The synthesized MEG signal $\hat{Y} = \mathcal{F}_\theta(X)$ should closely resemble the real MEG, while also capturing the lower spatial dispersion and higher high-frequency sensitivity characteristic of MEG compared to EEG.

\begin{figure*}[ht]
\centering
\includegraphics[width=0.97\linewidth]{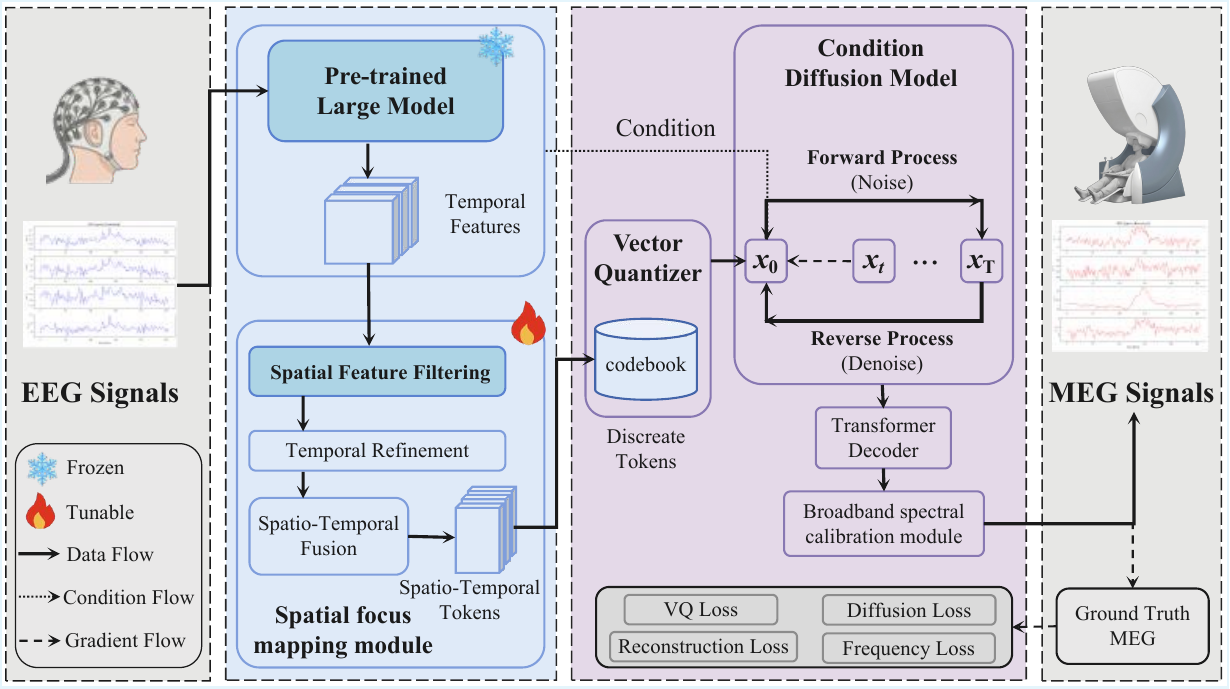}
\caption{Flowchart of the pre-trained EEG-to-MEG generative framework. The framework comprises four core components: a EEG representation encoder based on Pre-trained model to extract general neural activity representations. A spatial focus mapping module for EEG-MEG feature alignment. A latent conditional diffusion generator for latent space MEG feature generation. A broadband spectral calibration module for MEG signal recovery. The model is jointly optimized end-to-end via a multi-dimensional loss function.}
\label{fig:framework}
\end{figure*}

To learn this mapping, this paper proposes the generative framework shown in Fig.~\ref{fig:framework}. The framework consists of four core components: The EEG representation encoder based on a pre-trained model, which employs a channel-independent tokenization strategy to extract general neural activity representations from EEG data. The spatial focus mapping module, which aligns EEG and MEG in the latent source space through spatiotemporal attention and vector quantization. The latent conditional diffusion generator, which synthesizes MEG signals within a compressed latent space. The broadband spectral calibration module, which ensures the spectral fidelity of the synthesized MEG using frequency-domain modulation and multi-band supervision.

\subsection{EEG Representation Encoder based on Pre-trained Model}
To mitigate the impact of limited paired EEG-MEG data and leverage prior knowledge of generalizable EEG representations, we employ the pre-trained EEG model as the frozen backbone for feature extraction. Direct input of multi-channel data may introduce EEG-specific spatial dispersion, which hinders cross-modal alignment. We therefore implement a univariate sequence decomposition strategy to decouple temporal extraction from spatial transformation. Specifically, input EEG $X$ is reshaped into independent univariate sequences. This enables the pre-trained model to treat each EEG electrode as an independent observation unit, prioritizing the extraction of fine-grained channel-level temporal features while precluding premature spatial fusion.

Subsequently, these sequences are fed into a frozen pre-trained model. Here, we adopt LaBraM \cite{jiang2024large} due to its superior performance in capturing cross-subject neural patterns and compatibility with diverse electrode configurations. The EEG signals first pass through a patch embedding layer composed of 1D convolutions for tokenization, transforming them into patch sequences containing local waveform features. These sequences are propagated through deep transformer encoders to extract long-range temporal dependencies.

Finally, to adapt to the downstream EEG-to-MEG generation task, we perform spatial dimension restoration on the output features, restructuring them back into a tensor possessing channel topology. The entire universal temporal feature extraction process can be formalized as:
\begin{equation}
H_{enc} = \mathcal{R}_{out} \left( \mathcal{E}_{\text{pre}} \left( \mathcal{R}_{in}(X); \theta_{\text{fixed}} \right) \right) \in \mathbb{R}^{B \times C_e \times T' \times D},
\end{equation}
where $\mathcal{R}_{in}$ denotes the reshaping operation that decomposes multi-channel signals into univariate sequences, $\mathcal{E}_{\text{pre}}$ represents the frozen pre-trained encoder parameterized by $\theta_{\text{fixed}}$, and $\mathcal{R}_{out}$ denotes the inverse reshaping operation that restores the channel dimension. The resulting features $H_{enc}$ contain general semantic information and are passed as input to the subsequent tunable spatial focus mapping module.

\subsection{Spatial Focus Mapping Module}
Although EEG and MEG originate from the same neural activity, EEG exhibits significant spatial blurring due to the volume conduction effect of head tissues. This module aims to reorganize and focus features via spatiotemporal joint attention and align EEG with MEG in the latent source space through discrete projection.

First, to recover clear neural source features from diffuse EEG signals, we introduce spatial feature filtering based on self-attention. Specifically, we project the input features $H_{enc}$ using three independent learnable weight matrices $W_Q, W_K, W_V \in \mathbb{R}^{D \times D}$ to obtain query ($Q$), key ($K$), and value ($V$) matrices, respectively. Then, an affinity matrix between channels is computed to perform feature fusion in the spatial dimension:
\begin{equation}
H_{spa} = \text{Softmax}\left(\frac{Q K^T}{\sqrt{d_k}}\right) V + H_{enc},
\end{equation}
where $d_k$ is a scaling factor. Notably, this process is not merely feature aggregation. Instead, implicitly guided by the downstream MEG alignment objective, it dynamically suppresses diffuse background noise and redistributes channel importance. Consequently, it reorganizes the originally blurred channel features into a more spatially concentrated distribution.

Building on this, to capture the dynamic evolution of neural activity, we further apply attention mechanisms in the time dimension for spatiotemporal fusion. Unlike spatial attention which focuses solely on inter-channel relationships, this step aggregates context information across the entire time window, effectively capturing long-range temporal dependencies. This ensures that the model achieves not only spatial focusing but also precise synchronization with real neural oscillations in terms of temporal phase, thereby generating a more comprehensive spatiotemporal coupled representation.

Finally, to enforce the synthesized features to strictly conform to the MEG source space distribution, we introduce a Vector Quantizer. We construct a learnable codebook $\mathcal{E} = \{e_k\}_{k=1}^K$, where each codeword $e_k$ represents a typical MEG neural activation prototype. For each spatiotemporally fused feature vector $z^{(i)}$, we search for the nearest prototype in the codebook as its quantized representation $z_q^{(i)}$:
\begin{equation}
z_q^{(i)} = e_{k^*}, \quad \text{where } k^* = \mathop{\arg\min}_{k \in \{1,\dots,K\}} \| z^{(i)} - e_k \|_2.
\end{equation}

To optimize the codebook and ensure the encoder output approximates the MEG manifold, we define the VQ loss function $\mathcal{L}_{VQ}$ as follows:
\begin{equation}
\mathcal{L}_{VQ} = \| \text{sg}[H_{spa}] - Z_q \|_2^2 + \beta \| H_{spa} - \text{sg}[Z_q] \|_2^2,
\end{equation}
where $Z_q$ is the quantized feature tensor, $\text{sg}[\cdot]$ denotes the stop-gradient operator, and $\beta$ is the weight for the commitment loss. This step discretizes the continuous EEG features, achieving rigorous alignment with the MEG-specific manifold space.

\subsection{Latent Conditional Diffusion Generator}
We establish a probabilistic generative model in the quantized latent space to capture complex cross-modal dependencies. We treat the real MEG latent representation $z_0$ as the data distribution and gradually add Gaussian noise through a fixed Markov chain. At timestep $t$ of the forward process, the noisy state $z_t$ is defined as:
\begin{equation}
z_t = \sqrt{\bar{\alpha}_t} z_0 + \sqrt{1 - \bar{\alpha}_t} \epsilon, \quad \epsilon \sim \mathcal{N}(0, I),
\end{equation}
where $\epsilon$ is sampled standard Gaussian noise, and $\bar{\alpha}_t$ is the predefined noise variance schedule.

In the reverse generation phase, a denoising network $\epsilon_\theta$, which takes the current noisy state $z_t$, diffusion timestep $t$, and the aligned feature condition $c = Z_q$ from the previous module as input. The network's goal is to predict the added noise $\epsilon$. The optimization objective $\mathcal{L}_{diff}$ is defined as the mean squared error between the predicted noise and the true noise:
\begin{equation}
\mathcal{L}_{diff} = \mathbb{E}_{z_0, t, \epsilon, c} \left[ \| \epsilon - \epsilon_\theta(z_t, t, c) \|_2^2 \right].
\end{equation}

Through this process, the model learns to gradually recover latent temporal features conforming to the MEG distribution from pure noise, guided by the EEG semantic condition $c$.

\subsection{Broadband Spectral Calibration Module}
Due to the influence of head tissues, EEG signals typically suffer from high-frequency attenuation, whereas MEG preserves more broadband dynamic characteristics. To ensure that the synthesized MEG signals truthfully restore this property in the frequency domain, we design a calibration strategy comprising frequency domain modulation and multi-band supervision.

The broadband spectral calibration module begins with global frequency-domain modulation to correct spectral deviations. We introduce a learnable complex weight parameter $W_{freq}$ for frequency-domain global modulation. This parameter acts as an adaptive frequency-domain filter, modulating the spectrum of the synthesized signal $\hat{Y}_{raw}$ pointwise:
\begin{equation}
\hat{Y} = \text{IFFT}(\text{FFT}(\hat{Y}_{raw}) \odot W_{freq}),
\end{equation}
where $\odot$ denotes element-wise multiplication. Through end-to-end training, $W_{freq}$ automatically learns to adjust the gain and phase of each frequency component, enabling the spectral distribution of the output MEG $\hat{Y}$ to adaptively approximate the broadband characteristics of real MEG.

Complementing this, a band-specific loss is employed as the core spectral constraint $\mathcal{L}_{spec}$. We decompose the synthesized MEG signal $\hat{Y}$ and the real MEG signal $Y$ into five standard neurophysiological frequency bands: Delta, Theta, Alpha, Beta, and Gamma. For each specific band $b$, we compute the consistency error of its time-domain waveform:
\begin{equation}
\mathcal{L}_{spec} = \sum_{b \in \{\delta, \dots, \gamma\}} \left( \frac{1}{C_m T} \sum_{c, t} \| \hat{Y}_{b, c, t} - Y_{b, c, t} \|_2^2 \right),
\end{equation}
where $\hat{Y}_{b, c, t}$ and $Y_{b, c, t}$ represent the values of the synthesized and real signals in the $b$ band, respectively. This loss function forces the model to optimize both low-frequency fundamental rhythms and high-frequency rapid oscillations simultaneously. As a result, the synthesized MEG signals achieve consistency with real data in both time-domain waveforms and frequency-domain characteristics, effectively capturing broadband neural dynamics.

The framework adopts an end-to-end joint training strategy. The total loss function $\mathcal{L}_{total}$ is composed of four weighted terms:
\begin{equation}
\mathcal{L}_{total} = \lambda_{rec} \| Y - \hat{Y} \|_1 + \lambda_{vq} \mathcal{L}_{VQ} + \lambda_{diff} \mathcal{L}_{diff} + \lambda_{spec} \mathcal{L}_{spec},
\end{equation}
where $\| Y - \hat{Y} \|_1$ is the time-domain L1 reconstruction loss, ensuring consistency of waveform contours. And $\lambda_{rec}, \lambda_{vq}, \lambda_{diff}, \lambda_{spec}$ are the balancing coefficients for each loss term. This objective function constrains the model across multiple dimensions, including time-domain, frequency-domain, latent space distributions, and generative probability distributions, to ensure that the synthesized MEG signals possess both physical realism and physiological significance.

\section{Experiment}

\subsection{Datasets and Model Parameter Settings}

\begin{figure*}[ht]
\centering
\includegraphics[width=0.86\linewidth]{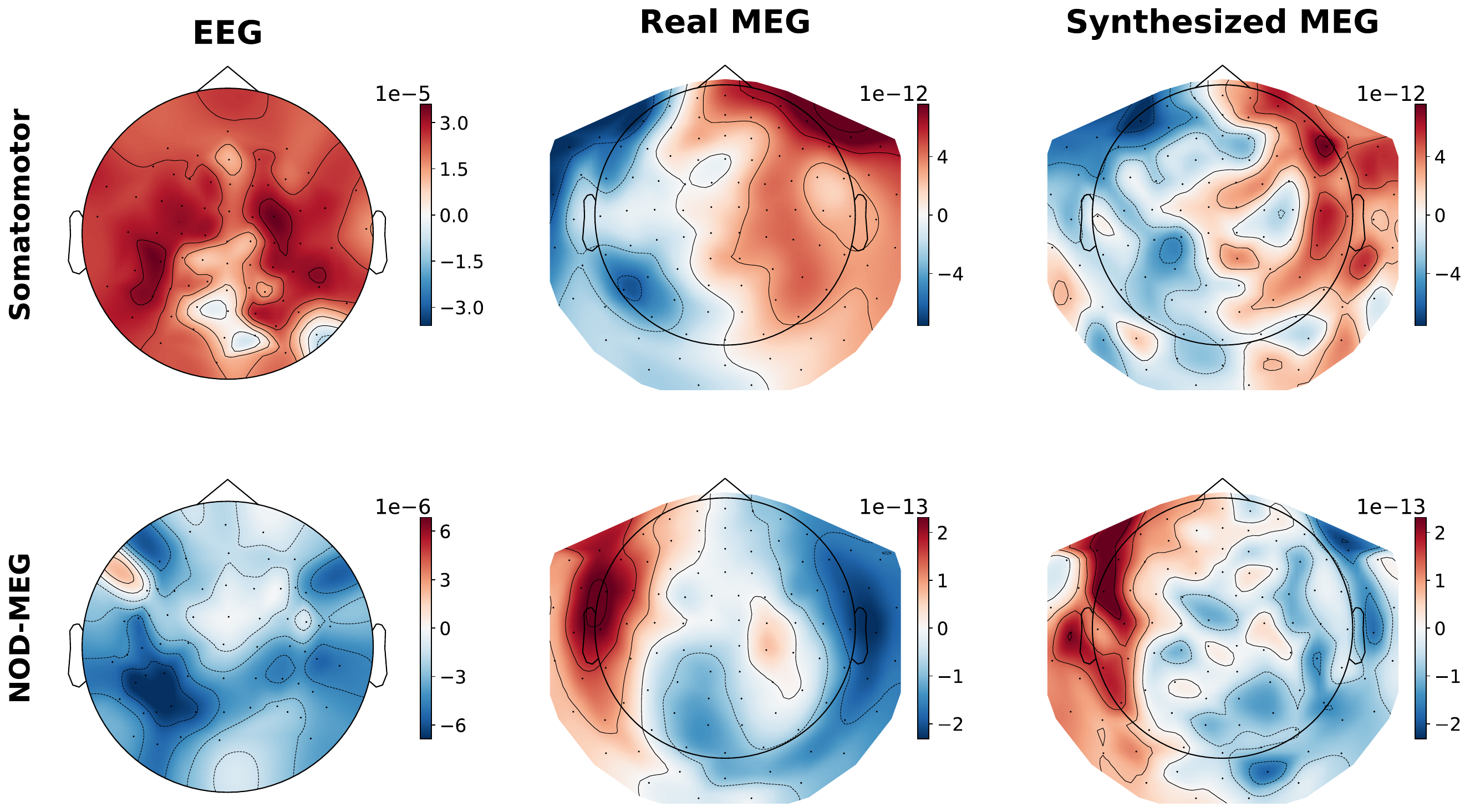}
\caption{Comparison of topographic maps from original EEG, real MEG, and synthesized MEG under task-activation states. The first row corresponds to the Somatomotor dataset, and the second row corresponds to the NOD-MEG dataset. The three columns display the input EEG, real MEG, and synthesized MEG topographies, respectively. The synthesized MEG topographies exhibit high spatial consistency with the real MEG, accurately capturing the task-specific activation patterns for both datasets.}
\label{fig:map}
\end{figure*}

\begin{figure*}[!t]
\centering
\includegraphics[width=0.93\linewidth]{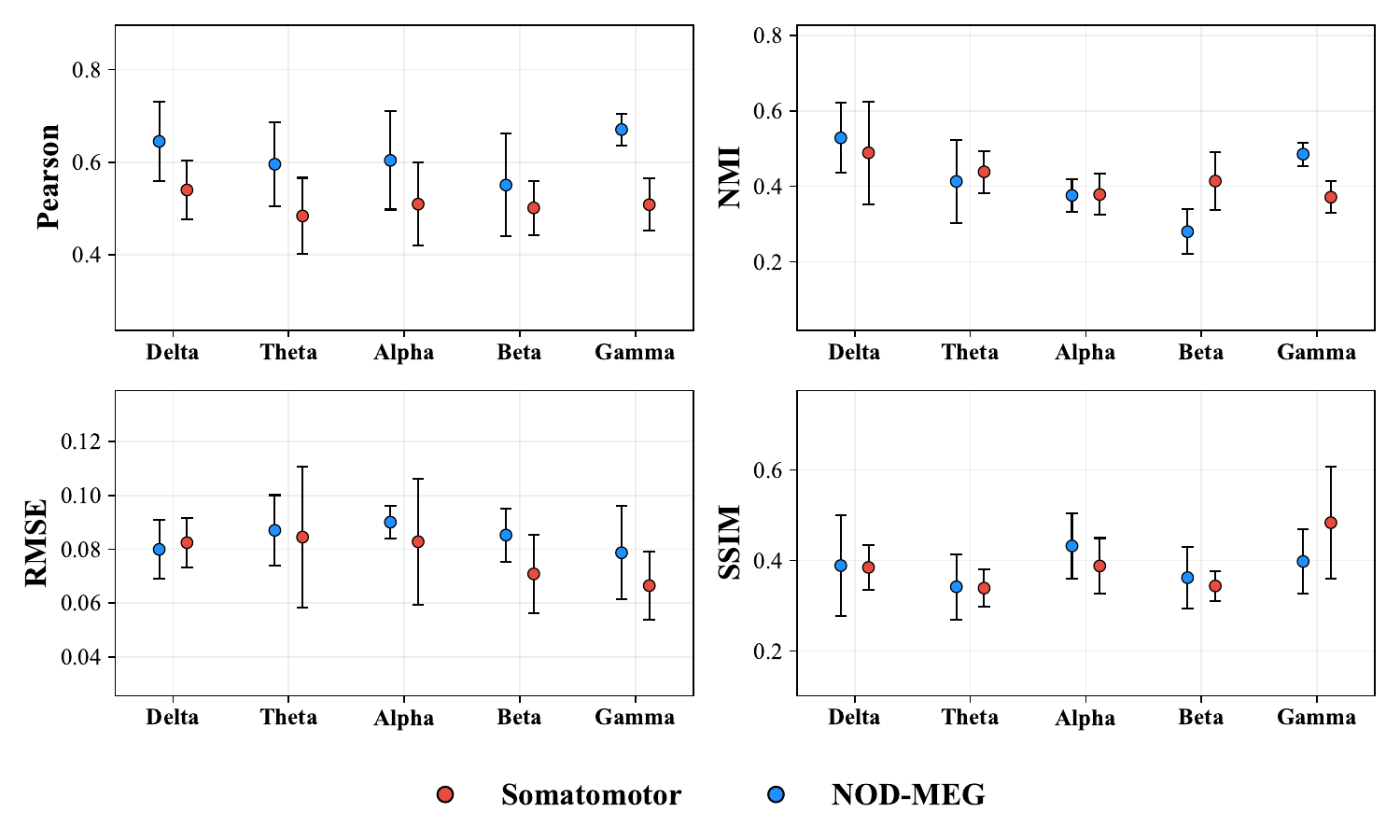}
\caption{Quantitative quality assessment of synthesized MEG across different frequency bands. This figure shows the comparison of four key performance metrics between synthesized and real MEG across five standard frequency bands (Delta, Theta, Alpha, Beta, Gamma). Red dots represent the Somatomotor dataset, while blue dots represent the NOD-MEG dataset. Data points indicate the mean values on the test set, and error bars denote the standard deviations. Quantitative analysis confirms that the synthesized MEG maintain high fidelity to real MEG across the full frequency spectrum, exhibiting consistent temporal alignment and spatial structural similarity.}
\label{fig:PCC}
\end{figure*}

\begin{figure*}[ht]
\centering
\includegraphics[width=0.92\linewidth]{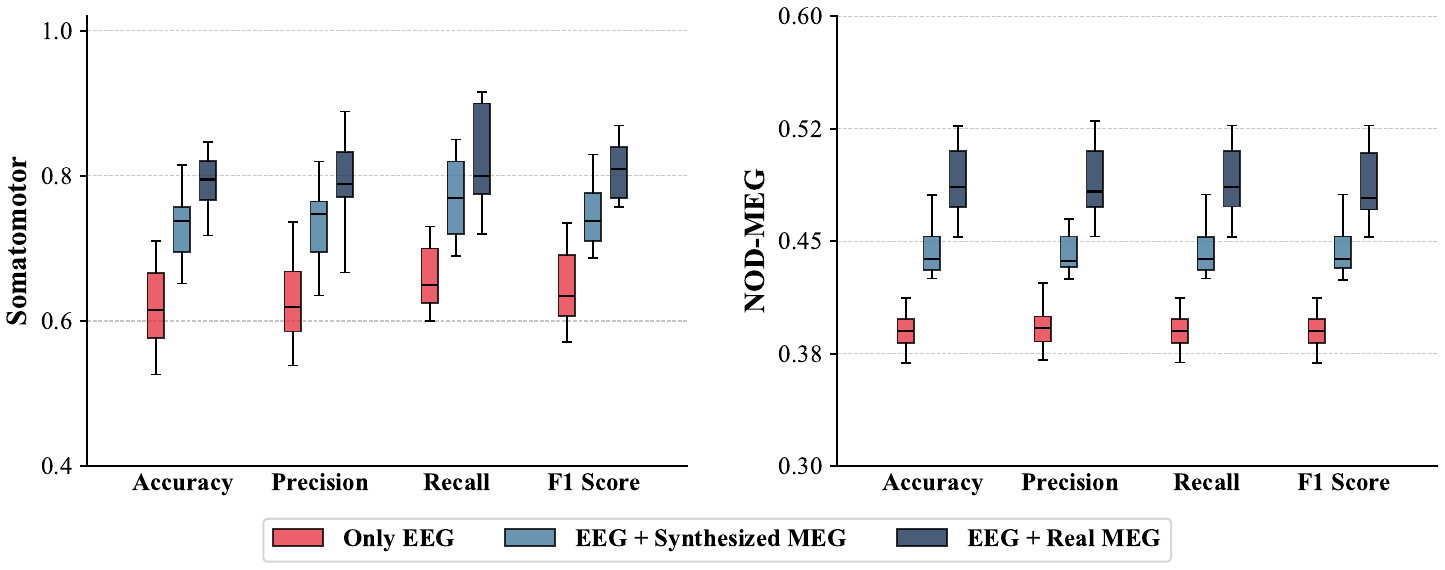}
\caption{Comparative assessment of BCI decoding performance. The left and right panels illustrate classification performance on the Somatomotor dataset and the NOD-MEG dataset, respectively. Three input modalities are compared: EEG only (red), EEG combined with synthesized MEG (light blue), and EEG combined with real MEG (dark blue). Box plots display the statistical distributions of accuracy, precision, recall, and F1 scores derived from ten runs of 10-fold cross-validation. The central horizontal line represents the median, and the upper and lower box edges indicate the quartiles. The results demonstrate that combining synthesized MEG and EEG significantly enhances decoding performance across both datasets compared to the only EEG baseline.}
\label{fig:class}
\end{figure*}

The core validation of this study was performed on two publicly available paired EEG and MEG datasets. All data were simultaneously acquired using the Elekta Neuromag system, ensuring strict temporal alignment between the EEG and MEG. The Somatomotor dataset\cite{ds006035:1.0.0} comprises a sensorimotor task, with the experimental paradigm involving stimulation of the median nerve at the wrist at regular intervals and requiring subjects to perform a finger lift as rapidly as possible after stimulation. The NOD-MEG dataset\cite{appelhoff2019mne} addresses a visual recognition task, with the experimental design entailing the repeated presentation of three distinct stimulus categories: famous faces, unfamiliar faces, and scrambled face images. Furthermore, to validate whether synthesized MEG can enhance BCI decoding performance in real-world BCI deployment where only EEG is available, we introduced additional only EEG datasets, including Illusory-face-eeg\cite{robinson2025neural}, EmoEEG-MC\cite{xu2025multi}, Kinesthetic motor imagery\cite{ds006940:1.0.0}, Motor imagery-Sarkar\cite{sarkar2024optimal}, and PerceiveImagine \cite{li2023multi} for supplementary testing.

The model is optimized using the AdamW optimizer \cite{loshchilovdecoupled} with hyperparameters: initial learning rate $\eta = 10^{-4}$, weight decay $\lambda_{\text{wd}} = 10^{-2}$, betas $\beta_1 = 0.9$ and $\beta_2 = 0.999$ (exponential decay rates for first and second moment estimates), and gradient clipping at maximum norm $\tau = 1.0$ to stabilize training and prevent gradient explosion. The model was trained on a single NVIDIA A800 GPU for 1000 epochs.

\subsection{MEG Generation Quality Assessment}

The visualization of EEG and MEG topographical maps is a method for evaluating the spatial distribution of neural activity, as it intuitively displays brain activation patterns and electromagnetic topological features during specific tasks. To visually assess the spatial consistency between the synthesized and real MEG, topographical maps of the EEG, synthesized MEG, and real MEG were compared under task-activation states. As shown in Fig.~\ref{fig:map}, the topographical maps of the synthesized MEG show a high consistency with the real MEG in terms of overall topological structure. In evaluations across different experimental paradigms, the framework demonstrates excellent spatial reconstruction capabilities, accurately recovering brain activation patterns associated with specific cognitive tasks. Although minor discrepancies exist in some local details, the high consistency in overall spatial distribution demonstrates that the framework has effectively captured the specific spatial characteristics of real MEG signals.

To quantitatively assess the spectral accuracy of the synthesized MEG signals, performance metrics were computed across five standard neurophysiological frequency bands. These five frequency bands are defined as: Delta (0.5–4 Hz), Theta (4–8 Hz), Alpha (8–13 Hz), Beta (13–30 Hz) and Gamma (30–100 Hz). Four evaluation metrics were employed: the Pearson correlation coefficient (Pearson) to assess temporal waveform consistency, and the Root Mean Square Error (RMSE) to evaluate the reconstruction error. Additionally, Normalized Mutual Information (NMI) and the Structural Similarity Index Measure (SSIM) were specifically employed to assess the similarity of the synthesized MEG signals in terms of spatial distribution and topological structure.

As shown in Fig.~\ref{fig:PCC}, quantitative analysis confirmed the similarity of the synthesized MEG signals to the real MEG across broadband frequencies. In terms of amplitude and temporal characteristics, RMSE values across all frequency bands remained below 0.1 for both the Somatomotor and NOD-MEG datasets, while Pearson coefficients exhibited positive correlations. Particularly in the Somatomotor dataset, Pearson coefficients ranged from 0.55±0.10 in the Beta band to 0.67±0.03 in the Gamma band, demonstrating consistent alignment between synthesized and real MEG signals across all frequency bands. Furthermore, regarding spatial structure, the SSIM and NMI metrics remained stable across all five frequency bands. Even within the Gamma band, which contains high-frequency information, the model maintained a high degree of structural similarity. These results show that the synthesized MEG signals successfully preserve the characteristics of the real MEG across the full bandwidth.

\begin{figure*}[ht]
\centering
\includegraphics[width=1\linewidth]{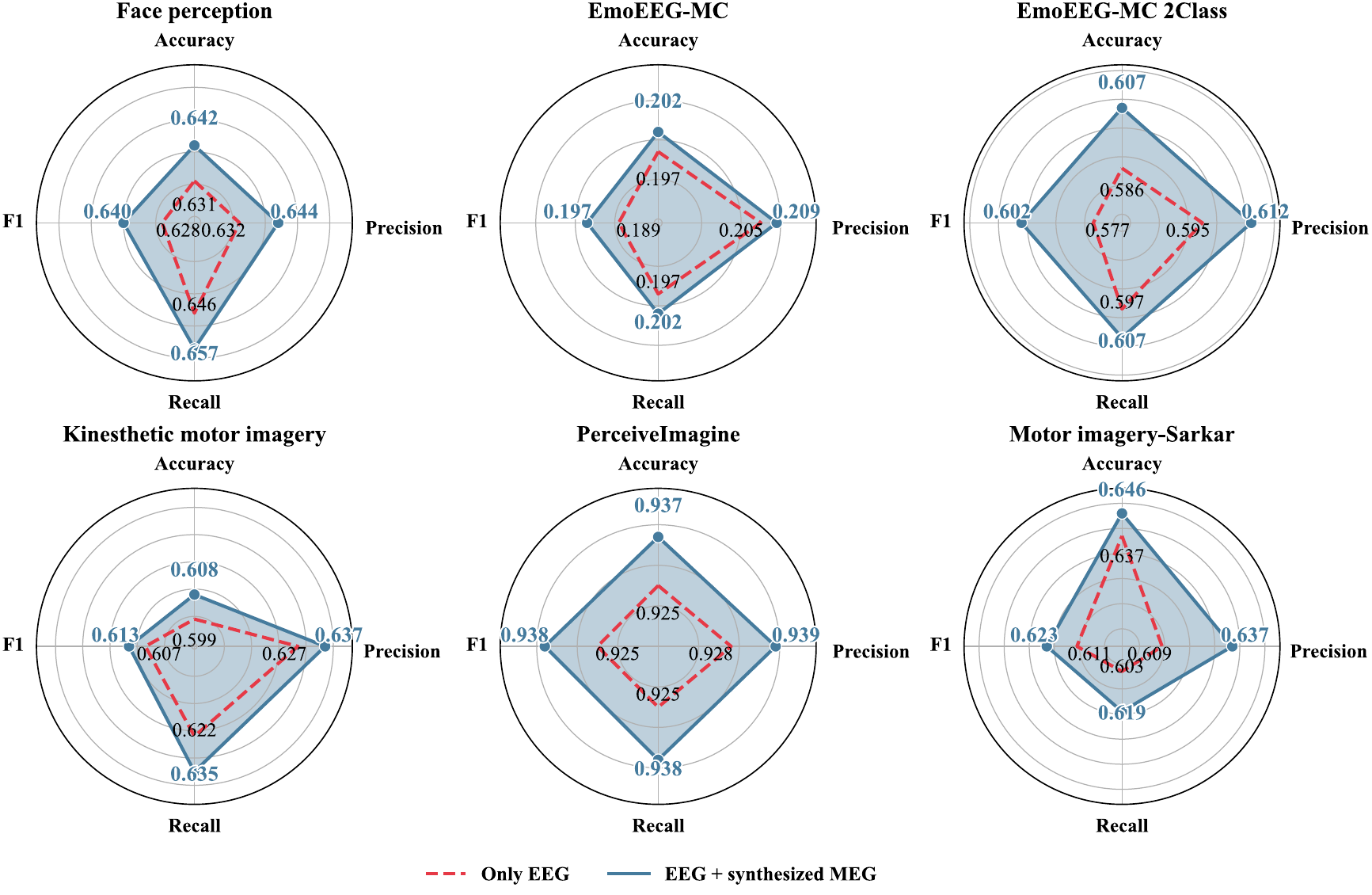}
\caption{Performance enhancement of BCI decoding using synthesized MEG on only EEG datasets. This figure shows the comparison of decoding performance before and after incorporating synthesized MEG signals across only EEG datasets. Radar charts display the mean scores for Accuracy, Precision, Recall, and F1 on the test sets. The red dashed line represents the baseline performance using EEG only, while the blue solid line and shaded area represent the performance achieved after combining synthesized MEG. The results demonstrate the potential of the proposed framework in cross-scenario BCI decoding.}
\label{fig:class2}
\end{figure*}

\subsection{BCI Decoding Performance Enhancement}

To validate the utility of synthesized MEG in practical BCI applications, we designed sensorimotor and visual perception decoding experiments based on paired EEG-MEG datasets. If the synthesized MEG only looks like the real MEG wave but has no neural information, it will not provide any benefit to decoding and might act as noise. To this end, a classic decoding pipeline combining feature engineering and support vector machines was adopted to simulate the operational environment of low-compute portable BCI systems. We employed 10 runs of 10-fold cross-validation for experiments on both the binary classification task of the Somatomotor dataset and the three-class classification task of the NOD-MEG dataset.

As shown in Fig.~\ref{fig:class}, on paired EEG-MEG datasets containing real MEG, multi-dimensional quantitative analysis confirmed that synthesized MEG significantly enhances BCI decoding performance. In the Somatomotor dataset of sensorimotor task, combining synthesized MEG and EEG improved the median decoding accuracy by 10.74\% compared to the EEG-only baseline. The F1 score increased from 0.6460 to 0.7259, indicating balanced improvements in both precision and recall. Similarly, in the NOD-MEG dataset of visual recognition task, the median decoding accuracy improved by 4.82\%, and the median F1 score increased from 0.3900 to 0.4381, validating the effectiveness of synthesized MEG signals in multi-class tasks. Crucially, the variance across evaluation metrics remained stable, indicating that the data augmentation did not introduce additional uncertainty. Although the performance of synthesized MEG is slightly below that of real MEG, its performance exceeding the EEG-only baseline demonstrates that it effectively overcomes the inherent limitations of single-modality EEG.

Furthermore, to evaluate the MEG synthesis model's performance across subjects and scenarios, we directly applied the trained MEG synthesis model to five completely independent datasets containing only EEG signals. These independent EEG datasets encompass distinct experimental paradigms, including face perception, emotion recognition, and motor imagery. Likewise, a support vector machine with feature engineering was utilized as the decoder, with results tested through 10 runs of 10-fold cross-validation.

\begin{figure*}[ht]
\centering
\includegraphics[width=0.69\linewidth]{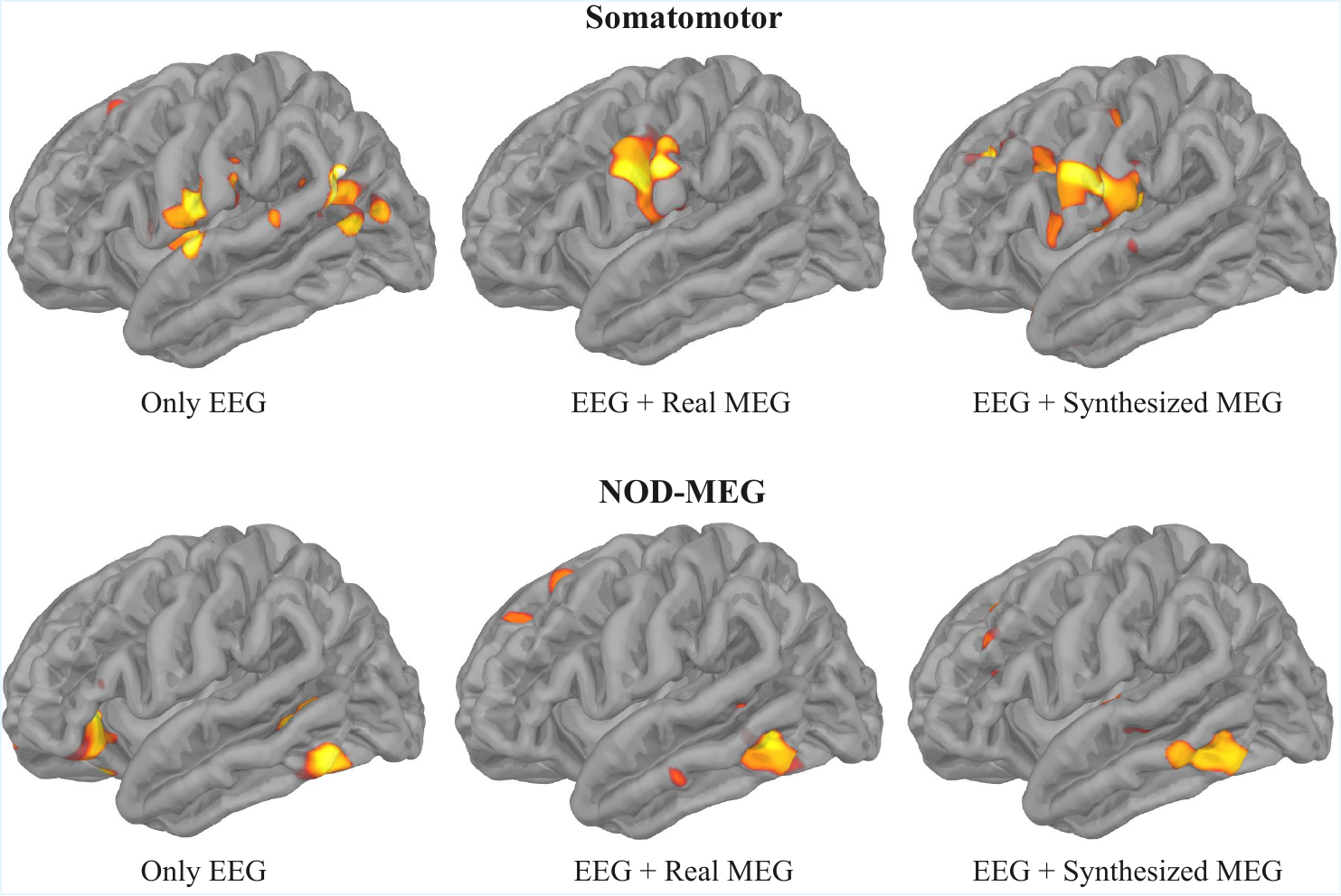}
\caption{Qualitative comparison of cortical source localization results. The figure depicts cortical source activation distributions across the Somatomotor and NOD-MEG datasets. The three columns correspond to three input conditions: Only EEG, EEG + Real MEG (serving as the reference standard), and EEG + Synthesized MEG. Results indicate that the cortical activations obtained by combining EEG with synthesized MEG are highly similar to the reference standard, significantly enhancing source imaging performance compared to using only EEG data.}
\label{fig:source}
\end{figure*}

As shown in Fig.~\ref{fig:class2}, we compared the decoding performance between the only EEG condition and the combining EEG with synthesized MEG condition. Quantitative results demonstrate that combining EEG with synthesized MEG improved decoding metrics across all only EEG datasets. For instance, in the PerceiveImagine dataset, accuracy further improved to 93.7\% from a high baseline of 92.5\%, while in the challenging EmoEEG-MC seven-class task, the F1 score increased from 0.189 to 0.197. This cross-task enhancement indicates that our framework successfully recovered MEG-specific attributes that are complementary to the original EEG. This proves that our framework captures general EEG-to-MEG mapping rules rather than merely fitting dataset-specific features. The synthesized MEG demonstrates significant potential for BCI applications across diverse scenarios.

\subsection{Source Imaging Analysis}
To further validate the synthesized MEG signals at the neurophysiological level, source imaging experiments were conducted. Dynamic Statistical Parametric Mapping (dSPM) was adopted, utilizing the fsaverage standard head model for source reconstruction. As shown in Fig.~\ref{fig:source}, source imaging results were compared across three conditions during task activation states. These conditions included: EEG only, EEG combined with real MEG, and EEG combined with synthesized MEG. The combination of EEG and real MEG serves as the reference standard here, as existing literature demonstrates that this combination significantly enhances imaging accuracy \cite{tong2024debiased, jiao2024multi}. Cortical activations were examined in the motor and visual cortices for the Somatomotor and NOD-MEG datasets, respectively.

In the Somatomotor dataset, source imaging results from combining EEG with real MEG showed highly focused spatial distribution, demonstrating the advantage of multimodal fusion in enhancing spatial resolution. In contrast, source imaging using only EEG exhibited significant spatial dispersion, with both blurred signal localization and reduced activation intensity. Notably, when EEG was combined with synthesized MEG, the spatial distribution pattern closely matched that of the real MEG condition. In the NOD-MEG dataset, source imaging results under all three conditions accurately localized to the face recognition region in the visual cortex. The introduction of synthesized MEG significantly improved imaging quality compared to the EEG-only condition. These results indicate that the synthesized MEG generated by our framework represents more than just sensor-level signal fitting. The synthesized MEG effectively approximates the cortical representation of real MEG at the source space level.

\section{Ablation Study}

To validate the impact of each module on MEG generation quality, we conducted ablation experiments on the Somatomotor dataset. As shown in Table \ref{tab:ablation}, where SFM module denotes the spatial focus mapping module and BSC module represents the broadband spectral calibration module, the decoding performance is optimal when combining EEG with the MEG synthesized by the full model. When the pre-trained encoder is removed, decoding performance significantly decreases, indicating that the general neural activity representations extracted by the pre-trained model are crucial for MEG generation. When the spatial focus mapping module and broadband spectral calibration module are removed, the performance of the decoding model also shows a significant decline. These results demonstrate the necessity of each module in constructing a high-quality EEG-to-MEG generation framework.

\begin{table}[h]
\centering
\caption{Ablation study results on decoding task.}
\label{tab:ablation}
\begin{tabular}{lcccc}
\toprule
\textbf{Modules} & \textbf{Accuracy} & \textbf{Precision} & \textbf{Recall} & \textbf{F1 Score} \\
\midrule
w/o Pre-trained encoder  & 0.6850 & 0.6900 & 0.7200 & 0.7047 \\
w/o SFM module & 0.6500 & 0.6450 & 0.6850 & 0.6644 \\
w/o BSC module & 0.6700 & 0.6750 & 0.7100 & 0.6921 \\
Full Model & \textbf{0.7086} & \textbf{0.7150} & \textbf{0.7467} & \textbf{0.7259} \\
\bottomrule
\end{tabular}
\end{table}

\section{Discussion and Conclusion}

This study proposes a cross-modal generation framework based on EEG-MEG spatiotemporal coupled representations to synthesize MEG signals. Comprehensive experiments confirm that the synthesized signals preserve signal characteristics at the sensor level and recover critical neural information in the source space. Furthermore, the framework significantly improved decoding performance in sensorimotor and visual recognition tasks. Importantly, performance enhancements were also achieved by combining synthesized MEG with EEG on independent EEG-only datasets. Collectively, these results demonstrate the feasibility of enhancing BCI decoding performance using synthesized MEG signals, highlighting their significant potential for practical BCI applications.

To further demonstrate the advantages of the proposed framework, we conducted a discussion comparing it with existing unpaired EEG-MEG learning paradigms\cite{xiao2025brainomni}. Although previous studies have attempted to leverage unpaired data for feature alignment between EEG and MEG, it is important to note that these studies did not truly achieve EEG-to-MEG generation. Fundamentally, EEG and MEG are synchronous electromagnetic manifestations of the same neural activity, sharing a strict spatiotemporal coupling relationship. Unpaired learning focuses solely on the similarity of global signal distributions, disrupting the temporal causality between electromagnetic signals at the millisecond timescale. Consequently, while such methods can simulate distributions consistent with the statistical characteristics of MEG, they fail to establish a deterministic mapping between specific EEG events and their corresponding MEG responses, nor can they guarantee phase performance. Therefore, employing strictly synchronized paired EEG-MEG data for supervised training is a necessary prerequisite for reconstructing accurate electromagnetic physical mappings and ensuring the clinical viability of synthesized MEG signals.

\begin{figure}[ht]
\centering
\includegraphics[width=\linewidth]{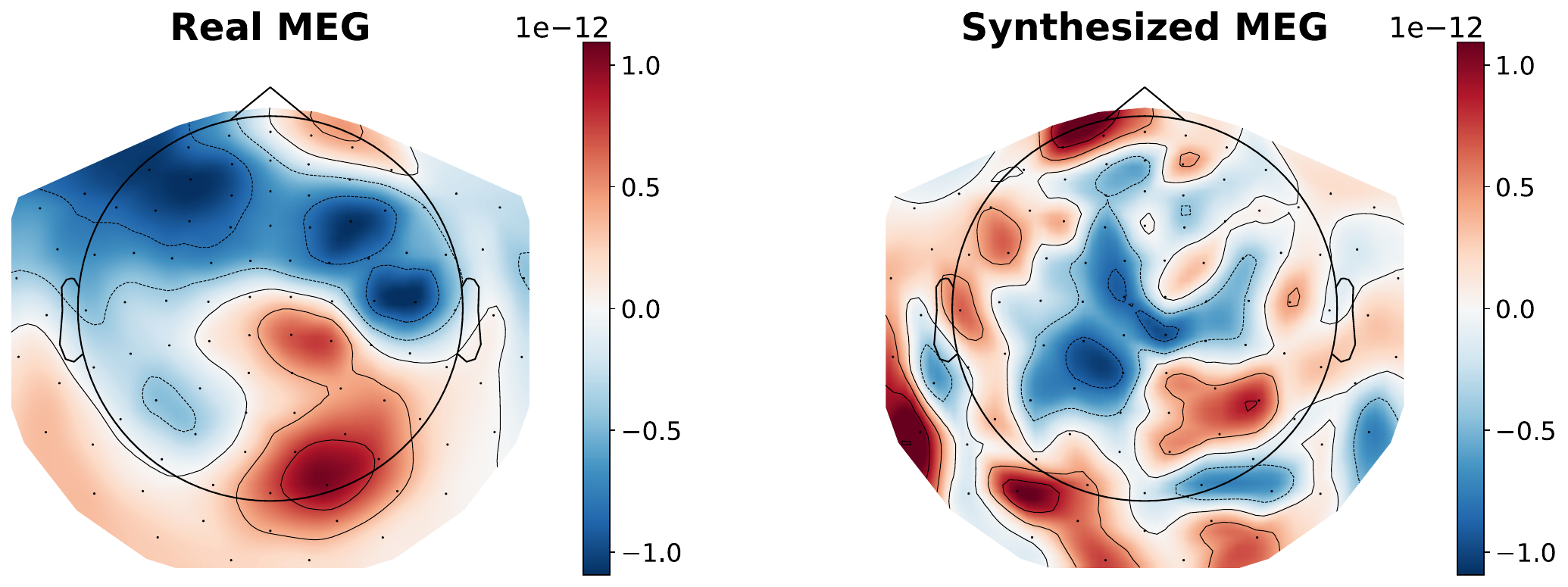}
\caption{Typical failure case of our method, showing significant differences between the real and synthesized MEG topographic maps.}
\label{fig:failed}
\end{figure}

Although this study develops a paired training-based framework for EEG-to-MEG generation, we acknowledge current limitations. Figure~\ref{fig:failed}  shows a typical failure case where synthesized MEG signals differ significantly from real MEG during neural activity. These discrepancies primarily result from two fundamental challenges. On one hand, the scarcity of paired EEG-MEG datasets constrains the model's capacity to learn how to generate MEG from EEG. On the other hand, MEG is a functional brain signal that exhibits significant individual variability. This cross-subject heterogeneity makes it difficult for the model to perfectly adapt to each subject's unique neurophysiological characteristics.

\begin{figure}[ht]
\centering
\includegraphics[width=\linewidth]{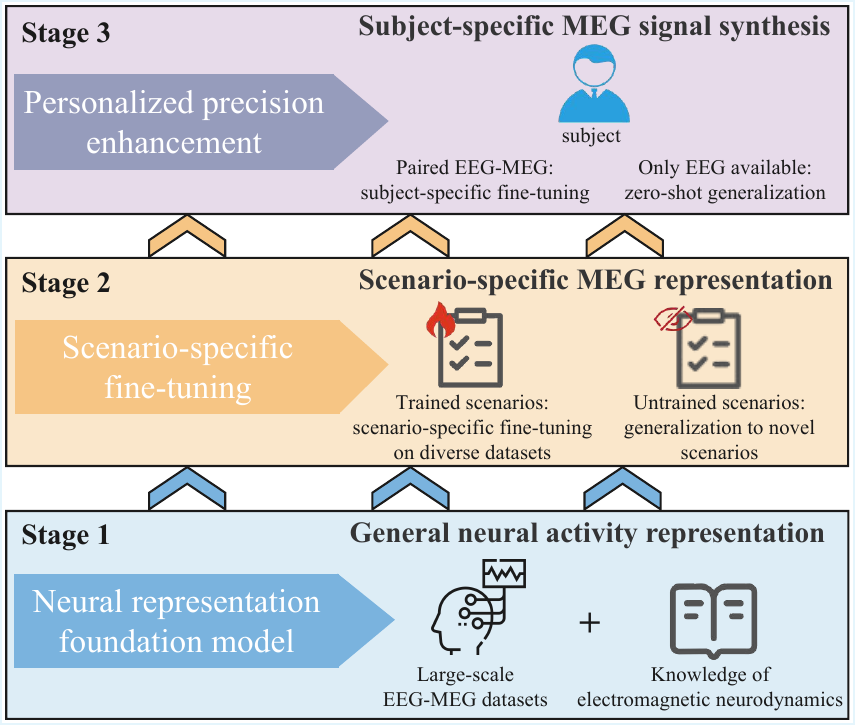}
\caption{The three-stage framework for future work. The bottom layer establishes a general foundation by integrating large-scale datasets with electromagnetic neural dynamics knowledge. The middle layer generates robust MEG representations, supporting fine-tuning on trained scenarios and generalization to novel scenarios. The top layer achieves personalized synthesis via subject-specific calibration or zero-shot generalization.}
\label{fig:future}
\end{figure}

To address these challenges, we propose a three-stage framework for future research as illustrated in Figure~\ref{fig:future}. First, we will build an EEG-MEG neural representation foundation model. By integrating large-scale EEG-MEG datasets with electromagnetic neural dynamics knowledge, the foundation model can effectively extract general neural activity representations. Second, we will focus on scenario-specific MEG representation. Using scenario-specific fine-tuning, the model will adapt to trained scenarios and generalize to new scenarios. Third, we will work on subject-specific MEG signal synthesis for personalized generation. For subjects with paired EEG-MEG data, we will use subject-specific fine-tuning. For subjects with only EEG data, we will use zero-shot generalization to synthesize MEG signals. This multi-stage framework will greatly advance non-invasive brain imaging and lay the foundation for low-cost and high-precision BCI.

\bibliographystyle{IEEEtran}
\bibliography{ref}

\end{document}